\documentstyle[pre,aps]{revtex}

\begin{document}

\draft

\title{Nonadiabatic Dynamics of Atoms in Nonuniform Magnetic 
Fields} 

\author{V.I. Yukalov} 

\address{International Centre of Condensed Matter Physics \\
University of Brasilia, CP 04513, Brasilia, DF 70919--970, Brazil\\
and\\
Bogolubov Laboratory of Theoretical Physics \\
Joint Institute for Nuclear Research, Dubna 141980, Russia \\
{\rm (Received: 2 January 1997)}}

\maketitle

\begin{abstract}

Dynamics of neutral atoms in nonuniform magnetic fields, typical of 
quadrupole magnetic traps, is considered by applying an accurate method for
solving nonlinear systems of differential equations. This method is more 
general than the adiabatic approximation and, thus, permits to check the 
limits of the latter and also to analyze nonadiabatic regimes of motion. 
An unusual nonadiabatic regime is found when atoms are confined from one 
side of the $\;z\;$--axis but are not confined from another side. The 
lifetime of atoms in a trap in this semi--confining regime can be 
sufficiently long for accomplishing experiments with a cloud of such 
atoms. At low temperature, the cloud is ellipsoidal being stretched in 
the axial direction and moving along the $\;z\;$--axis. The 
possibility of employing the semi--confining regime for studying the
relative motion of one component through another, in a binary mixture of 
gases is discussed.
\end{abstract}


\section{Introduction}

The motion of neutral atoms in nonuniform magnetic fields is important to 
study for several applications, in particular, for better understanding 
the mechanics of confinement in quadrupole magnetic traps, such as the
Ioffe--Pritchard traps with a static bottle field [1-3] or dynamic traps 
with a rotating bias field [4,5]. This has become especially interesting 
after the experimental observation of Bose condensation in very cold 
gases of Rubidium [6,7], Lithium [8], and Sodium [9-11]. There exists
extensive literature considering statistical properties of confined Bose 
systems, using various approaches, e.g. the quasiclassical 
density--of--state approximation [12,13], the Gross--Pitaevskii equation
[14-16], the Monte Carlo density--matrix calculations [17], the 
Thomas--Fermi approximation [18,19], the Bogolubov approximation [20], 
and the gas approximation in the frame of the Gibbs ensembles [21,22]. 
Statistical properties of the weakly interacting Fermi gas confined in a 
potential well have also been studied in the Thomas--Fermi approximation 
[23,24]. 

The aim of this paper is to consider not statistics but dynamics of atoms 
in nonuniform magnetic fields. When one is interested in the 
behaviour of confined atoms, one deals with their stationary motion. 
Stationary regimes can be described by the adiabatic approximation. 
When the confining potential is harmonic, then the dynamics of atoms is 
given by simple harmonic oscillations. In general, the confining 
potential is not necessarily harmonic. For instance, the first demonstrated 
magnetic trap [25] used a quadrupole field (with zero magnetic field at 
the center) which gave rise to a linear potential. In any case, atomic 
motion in a strictly confining potential can be described by the 
adiabatic approximation. Such adiabatic motion in various magnetic traps 
has been analysed in Refs. [26-29].

A more general consideration of atomic motion, without using the 
adiabatic approximation, is meaningful for several reasons: First
of all, a more general approach makes it possible to understand the 
limits of the adiabatic approximation. Second, studying other, 
nonadiabatic, regimes of motion permits to explain more profoundly the
physics of atoms inside magnetic traps, as far as in these not all atoms 
are confined. Knowing better different, including nonadiabatic, regimes 
of atomic motion may, possibly, give a hint on how to improve confining 
characteristics of magnetic traps.

One more reason is related to the recently reported experiments on the 
simultaneous trapping of two different atomic species, two isotopes of 
rubidium [30] and  sodium and potassium [31]. These experiments are a starting 
point for a new series of studies of ultracold matter. The variety of effects 
that can be observed in mixtures are incomparably richer than in 
one--component gases. This concerns even equilibrium mixtures [32]. Much more 
interesting features appear when one of the components can move through 
another. For instance, in a binary mixture with such a relative motion the 
effect of conical stratification [33] can occur. This effect happens when one 
of the components moves in one direction through another component. Then the 
instability can develop inside a cone with the axis along the relative 
velocity; as a result of this instability, the components stratify in 
space. Note that for this effect the one--directional relative motion 
is necessary, but not relative oscillations or collisions of sloshing clouds.
In the presence of a relative macroscopic motion some unusual manifestations
of the Doppler effect [34] may also arise. But can such a strange regime 
exist when the atoms of one kind move in one direction being confined 
from another? In addition, it is desirable that this one--directional escape 
from a trap would not be too fast in order to be able to accomplish 
measurements. These requirements look too severe to allow the existence of 
such a semi--confining regime. However, in what follows it will be shown 
that this semi--confining regime does exist. Certainly, it is nonadiabatic 
and, even more, nonpotential, thus cannot be described in the frame of 
the adiabatic approximation.

\section{Evolution Equations}

Since the aim of this paper is to present an accurate solution of 
evolution equations for atoms in nonuniform magnetic fields, it is 
reasonable, first of all, to pay attention to the accurate formulation of 
the equations themselves.

The quantum Hamiltonian of a system of $\;N\;$ neutral atoms, each with a 
mass $\;m\;$ and magnetic moment $\;\mu\;$, is
\begin{equation}
H=\sum_{i=1}^N\left (\frac{\stackrel{\rightarrow}{p}_i^2}{2m} -\mu
\stackrel{\rightarrow}{S}_i\stackrel{\rightarrow}{B}_i\right ) +
\frac{1}{2}\sum_{i\neq j}^N\Phi_{ij} ,
\end{equation}
where $\;\stackrel{\rightarrow}{p}_i\;$ is a momentum operator; 
$\;\stackrel{\rightarrow}{S}_i\;$, a spin operator; 
$\;\stackrel{\rightarrow}{B}_i=
\stackrel{\rightarrow}{B}(\stackrel{\rightarrow}{r}_i,t)\;$ is the total 
magnetic field acting on an $\;i\;$--atom; and $\;\Phi_{ij}=
\Phi(|\stackrel{\rightarrow}{r}_i-\stackrel{\rightarrow}{r}_j|)\;$ is an 
interaction potential. The magnetic moment $\;\mu\;$ is the product of 
the Bohr magneton and the hyperfine $\;g\;$--factor [35]. The wave 
function of the system, $\;\Psi = 
[\Psi_s(\stackrel{\rightarrow}{r}_1,\stackrel{\rightarrow}{r}_2,\ldots ,
\stackrel{\rightarrow}{r}_N,t)]\;$ is a column in spin space. The 
quantum--mechanical average of an operator from the algebra of 
observables, $\;{\cal A}\;$, is the scalar product
\begin{equation}
\langle{\cal A}\rangle = (\Psi,{\cal A}\Psi) .
\end{equation}

Using the Schr\"odinger equation $\;i\hbar\partial\Psi/\partial 
t=H\Psi\;$, with the Hamiltonian (1), it is straightforward to get the 
evolution equations for the average position of an atom,
\begin{equation}
\frac{d}{dt}\langle\stackrel{\rightarrow}{r}_i\rangle =\frac{1}{m}\langle
\stackrel{\rightarrow}{p}_i\rangle ,
\end{equation}
its average momentum
\begin{equation}
\frac{d}{dt}\langle\stackrel{\rightarrow}{p}_i\rangle =\mu\langle
\stackrel{\rightarrow}{\nabla}_i(\stackrel{\rightarrow}{S}_i
\stackrel{\rightarrow}{B}_i)\rangle - \sum_{j(\neq i)}^N\langle
\stackrel{\rightarrow}{\nabla}_i\Phi_{ij}\rangle ,
\end{equation}
and the average spin
\begin{equation}
\frac{d}{dt}\langle\stackrel{\rightarrow}{S}_i\rangle =
\frac{\mu}{\hbar}\langle\stackrel{\rightarrow}{S}_i\times
\stackrel{\rightarrow}{B}_i\rangle .
\end{equation}

These equations may be simplified with a mean--field approximation
\begin{equation}
\langle S_i^\alpha B_i^\beta\rangle =\langle S_i^\alpha\rangle
\langle B_i^\beta\rangle ,
\end{equation}
where $\;\alpha,\beta=x,y,z\;$, valid for fields slowly varying in space 
[35]. Condition (6) is also called the semi--classical approximation and 
is usually supplemented by another approximate equation
\begin{equation}
\langle\stackrel{\rightarrow}{B}(\stackrel{\rightarrow}{r}_i,t)\rangle =
\stackrel{\rightarrow}{B}(\langle\stackrel{\rightarrow}{r}_i\rangle,t)
\end{equation}
which again assumes a slow variance of magnetic field in space. When the 
magnetic field is a linear function of real--space coordinates, then Eq.(7) 
is not an approximation but an exact relation. This concerns quadrupole 
magnetic fields which in what follows we shall deal with. Hence, the sole 
approximation we need is the mean--field one Eq.(6).

Under conditions (6) and (7), equations (3)--(5) acquire the same form for 
all indices $\;i\;$, which permits us to simplify the notation by introducing
$$ \stackrel{\rightarrow}{r} \equiv \langle\stackrel{\rightarrow}{r}_i
\rangle = \{ x,y,z\} , $$
\begin{equation}
\stackrel{\rightarrow}{v} \equiv \frac{1}{m}\langle
\stackrel{\rightarrow}{p}_i\rangle = \{ v_x,v_y,v_z\} ,
\end{equation}
$$ \stackrel{\rightarrow}{S} \equiv \langle
\stackrel{\rightarrow}{S}_i\rangle =\{ S_x,S_y,S_z\} , $$
and the average interatomic force
\begin{equation}
\stackrel{\rightarrow}{f}\equiv -\sum_{j(\neq i)}^N\langle
\stackrel{\rightarrow}{\nabla}\Phi_{ij}\rangle .
\end{equation}
Then Eqs.(3)--(5) can be reduced to the system of equations
$$ \frac{d\stackrel{\rightarrow}{r}}{dt} =\stackrel{\rightarrow}{v} , $$
\begin{equation}
\frac{d\stackrel{\rightarrow}{v}}{dt} =\frac{\mu}{m}
\stackrel{\rightarrow}{\nabla}(\stackrel{\rightarrow}{S}
\stackrel{\rightarrow}{B}) + \frac{\stackrel{\rightarrow}{f}}{m} ,
\end{equation}
$$ \frac{d\stackrel{\rightarrow}{S}}{dt} =\frac{\mu}{\hbar}
\stackrel{\rightarrow}{S}\times \stackrel{\rightarrow}{B} , $$
which will be the main object of our consideration.

The derivation of (10), though simple enough, contains an important point 
which is worth emphasizing. The basic approximation (6) supposes that the 
system fields slowly vary in real space, so that the spin and real--space 
degrees of freedom can be separated, which in the quantum--mechanical 
language means that the wave function can be factorized into a product of 
spin and real--space wave functions [35]. This quantum--mechanical 
separation of variables, as will be shown in what follows, is closely 
related to their dynamical separation.

The evolution equations (10) are to be supplemented by the initial 
conditions
$$ \stackrel{\rightarrow}{r}(0)=
\stackrel{\rightarrow}{r}_0 =\{ x_0,y_0,z_0\} , $$
\begin{equation}
\stackrel{\rightarrow}{v}(0) =\stackrel{\rightarrow}{v}_0 =
\{ v_0^x,v_0^y,v_0^z\} ,
\end{equation}
$$ \stackrel{\rightarrow}{S}(0) =\stackrel{\rightarrow}{S}_0 =
\{ S_0^x,S_0^y,S_0^z\} . $$

In specifying the form of the magnetic field, let us take it as in the 
experiments [5-7] with dynamical quadrupole traps. Then the total 
magnetic field
\begin{equation} 
\stackrel{\rightarrow}{B}= \stackrel{\rightarrow}{B}_1
(\stackrel{\rightarrow}{r}) +\stackrel{\rightarrow}{B}_2(t)
\end{equation}
is the sum of the quadrupole field
\begin{equation}
\stackrel{\rightarrow}{B}_1
(\stackrel{\rightarrow}{r}) = B_1'
(\stackrel{\rightarrow}{r} -3z\stackrel{\rightarrow}{e}_z)
\end{equation}
and the rotating bias field
\begin{equation}
\stackrel{\rightarrow}{B}_2(t) = B_2(
\stackrel{\rightarrow}{e}_x\cos \omega t +
\stackrel{\rightarrow}{e}_y\sin\omega t ) ,
\end{equation}
where $\;\stackrel{\rightarrow}{e}_\alpha\;$ is a unit vector for 
$\;\alpha=x,y,z\;$.

The characteristic length
\begin{equation}
L\equiv\frac{B_2}{B_1'}
\end{equation}
of the quadrupole--field nonuniformity corresponds to the radius of the 
field zero in the radial direction. This length defines approximately the 
upper limit for the radius of a trapped atomic cloud. Keeping this in 
mind, it is convenient to pass to dimensionless space variables measuring 
the components of the Cartesian vector $\;\stackrel{\rightarrow}{r}=
\{ x,y,z\}\;$ in units of $\;L\;$. Then we can make profit from the 
inequality
\begin{equation}
|\stackrel{\rightarrow}{r}| < 1 .
\end{equation}
To return to dimensional space variables, we have to put
$\;\stackrel{\rightarrow}{r} \rightarrow \stackrel{\rightarrow}{r} L\;$.

We introduce the characteristic frequencies
\begin{equation}
\omega_1\equiv\left (\frac{\mu B_1'}{mL}\right )^{1/2} , \qquad
\omega_2\equiv\frac{\mu B_2}{\hbar}
\end{equation}
of atomic and spin motions, respectively, and the collision rate 
$\;\gamma\;$ defined by the ratio
\begin{equation}
\gamma\stackrel{\rightarrow}{\xi}\equiv
\frac{\stackrel{\rightarrow}{f}}{mL} ,
\end{equation}
in which $\stackrel{\rightarrow}{\xi}$ is treated as a stochastic 
variable representing the interactions of atoms through their random 
collisions. A detailed definition of the variable 
$\stackrel{\rightarrow}{\xi}$ will be given in Sec.V and in Appendix.
Then from Eq.(10) we obtain the evolution equations for the space variable
\begin{equation}
\frac{d^2\stackrel{\rightarrow}{r}}{dt^2} =\omega_1^2\left ( S_x
\stackrel{\rightarrow}{e}_x +S_y\stackrel{\rightarrow}{e}_y -
2S_z\stackrel{\rightarrow}{e}_z\right ) +
\gamma\stackrel{\rightarrow}{\xi}
\end{equation}
and for the spin variable
\begin{equation}
\frac{d\stackrel{\rightarrow}{S}}{dt} =\omega_2\hat A
\stackrel{\rightarrow}{S} ,
\end{equation}
where the matrix $\;\hat A=[A_{\alpha\beta}]\;$, with 
$\;\alpha,\beta=1,2,3\;$, consists of the elements
$$ A_{11}=A_{22}=A_{33}=0 , $$
$$ A_{12}=-A_{21}=-2z , $$
$$ A_{13}=-A_{31}=-y-\sin\omega t , $$
$$ A_{23} =-A_{32} = x+\cos\omega t . $$

We are concerned about an accurate solution of the system (19) and (20) of
nonlinear differential equations, without using the adiabatic 
approximation. At this point, to distinguish what is what, it is useful 
to say several words about the adiabatic approximation. Fortunately, this 
will not take too much space, since the latter approximation is rather 
trivial. First, one assumes that the dynamical process is close to its 
stationary state, so that it is admissible to put 
$\;d\stackrel{\rightarrow}{S}/dt=0\;$. Then, from the third equation 
in (10) it follows that $\;\stackrel{\rightarrow}{S}\times
\stackrel{\rightarrow}{B}=0\;$. This means that the spin is aligned along 
$\;\stackrel{\rightarrow}{B}\;$, which can be written as
$\;\stackrel{\rightarrow}{S}\cdot\stackrel{\rightarrow}{B}=
S|\stackrel{\rightarrow}{B}|\;$. Thus, one excludes the spin motion saying 
that spin adiabatically follows the magnetic field. Using this, for the 
atomic space variable $\;\stackrel{\rightarrow}{r}\;$ one gets the Newton
equation with the adiabatic force
$$ \stackrel{\rightarrow}{F}_a = S\mu
\stackrel{\rightarrow}{\nabla}|\stackrel{\rightarrow}{B}| . $$
Assume that the field rotates much faster than the mechanical 
oscillations of atoms, but not so fast as to induce transitions in the 
Zeeman substructure. This means that $\;\omega_1\ll\omega\ll\omega_2\;$. 
Then the adiabatic force can be averaged over the period $\;2\pi/\omega\;$ 
of the rotating bias field. Averaging this force and using Eq.(16), one gets
$$ \langle \stackrel{\rightarrow}{F}_a\rangle_t \simeq\frac{m}{2}S\omega_1^2
(x\stackrel{\rightarrow}{e}_x +y\stackrel{\rightarrow}{e}_y +
8z\stackrel{\rightarrow}{e}_z) , $$
which immediately yields the adiabatic potential
$$ U_a \simeq -\frac{m}{4}S\omega_1^2(x^2+y^2+8z^2) + const \; . $$
This is a harmonic, although anisotropic, potential. The motion of atoms 
in such a potential is given by simple harmonic oscillations, if $\;S<0\;$.
When $\;S=0\;$, atoms fly away ballistically and if $\;S>0\;$, they 
escape by the exponential law. In both latter cases atoms escape from the 
trap in all directions. The ballistic flying away is isotropic. The 
exponential escape, because of the anisotropy of the adiabatic 
potential, is anisotropic: atoms escape faster along the axial direction 
than in the radial one; but anyway the symmetry with respect to the 
inversion $\;\stackrel{\rightarrow}{r}\rightarrow -\stackrel{\rightarrow}{r}\;$
is preserved.
 
\section{Scale Separation}

Return to the general equations (19) and (20). Written in the standard 
form, they compose a nonlinear dynamical system of the ninth order, that 
is, a system of nine nonlinear differential equations. It seems that it 
is impossible to solve this complicated system of equations without 
invoking a rough approximation like the adiabatic one. Nevertheless, 
these equations can be solved using the method of scale separation [36,37]. 
The mathematical foundation of this approach is based on the Krylov--Bogolubov
averaging method [38] and the Poincar\'e theory of generalized asymptotic 
expansions [39]. The method of scale separation was successfully applied to
several intricate problems, such as the origin of self--organized 
spin superradiance in nuclear magnets [36], coherent radiation regimes of spin
masers [40,41], and fast polarization reversal in proton targets used for 
studying beam scattering [42]. The accuracy of this approach has been 
confirmed by good agreement of its solutions with experimental data [43,44] 
and with computer simulations [45,46].

The first step of the method of scale separation [36,37] is to classify 
the functional variables of the problem, separating time scales. Fortunately, 
in the majority of interesting physical problems it is possible to separate 
relatively slow from relatively fast variables. In our case, such a separation
is naturally related to the mean--field condition (6) used for deriving the 
evolution equations (19) and (20). As has been already discussed above, 
condition (6) assumes that the real--space nonuniformity in the system is, in 
some sense, small. Now we ascribe an exact meaning to this phrase concretizing
in what sense the nonuniformity is small. The nonuniformity in the system
is connected with the quadrupole field (13) and pair interactions in (9),
while the rotating bias field (14) is spatially uniform. Therefore, what we 
need is to compare the characteristic parameters related to the corresponding 
fields. Among these characteristic parameters we have $\;\omega_1\;$ and 
$\;\omega_2\;$ in Eq.(17), and $\;\gamma\;$ in Eq.(18). The nonuniformity is 
weak if the characteristic parameters corresponding to nonuniform fields are 
small as compared to that of a uniform field. The latter means nothing but the 
validity of the inequalities
\begin{equation}
\omega_1\ll\omega_2 , \qquad \gamma\ll\omega_2 .
\end{equation}

The meaning of the first inequality in (21) is quite evident, indicating that 
the frequency of mechanical oscillations of atoms is much smaller than 
that of spin fluctuations. The second inequality is also very natural, as 
far as the collision rate is usually much smaller than $\;\omega_2\;$. If 
$\;\gamma\;$ would be comparable with $\;\omega_2\;$ this would imply 
atomic collision are causing radio--frequency transitions between 
magnetic sublevels. With Eq.(21) in mind, looking at the evolution 
equations (19) and (20), we  notice at once that the variable 
$\;\stackrel{\rightarrow}{r}\;$ is to be treated as slow, compared to the 
fast variable $\;\stackrel{\rightarrow}{S}\;$.

At this point, it is worth emphasizing how naturally the separation of 
functional variables into slow and fast, with respect to time, is connected 
with the character of nonuniformity in real space. This is why spending 
some time for remembering the derivation of Eqs.(10) was not in vain, 
but, on the contrary, is important for stressing the self--consistency of 
the approximations used.

Following further the method of scale separation [36,37], we need to 
solve the equations for fast variables, with slow variables being kept as 
quasi--integrals of motion. The evolution equation for fast variables is
Eq.(20) for spin. This equation for an arbitrary given antisymmetric 
matrix $\;\hat A\;$, with elements $\;A_{ij}=-A_{ji}\;$, can be solved 
exactly. This means that we are able to present an exact solution for 
$\;\stackrel{\rightarrow}{S}\;$ for any given external fields. Because 
of the significance of such s solution, we write it down explicitly.

First, we solve the eigenproblem
\begin{equation}
\hat A\stackrel{\rightarrow}{b}_i=\alpha_i\stackrel{\rightarrow}{b}_i ,
\qquad |\stackrel{\rightarrow}{b}_i|^2 = 1 ,
\end{equation}
in which $\;\hat A\;$ is an antisymmetric matrix and $\;i=1,2,3\;$. The 
solution is straightforward giving the eigenvalues
\begin{equation}
\alpha_1=i\alpha , \qquad \alpha_2 =-i\alpha , \qquad \alpha_3 = 0 ,
\end{equation}
with
$$ \alpha\equiv\sqrt{A_{12}^2+A_{13}^2+A_{23}^2} . $$
The eigenvectors are
\begin{equation}
\stackrel{\rightarrow}{b}_i =\frac{1}{\sqrt{C_i}}\left [\left 
(\alpha_iA_{13}+A_{12}A_{23}\right) \stackrel{\rightarrow}{e}_x +
\left (\alpha_iA_{23}-A_{12}A_{13}\right)\stackrel{\rightarrow}{e}_y +
\left (\alpha_i^2+A_{12}^2\right )\stackrel{\rightarrow}{e}_z\right ]
\end{equation}
with the normalization constant
$$ C_i=\left (|\alpha_i|^2 -A_{12}^2\right )^2 + 
\left (|\alpha_i|^2+A_{12}^2\right )\left ( A_{13}^2 +A_{23}^2\right ) . $$
It can be checked straightaway that the vectors from Eq.(24) form an 
orthonormal basis and satisfy the properties
\begin{equation}
\stackrel{\rightarrow}{b}_i^*\stackrel{\rightarrow}{b}_j =\delta_{ij} ,
\qquad \stackrel{\rightarrow}{b}_1^*= \stackrel{\rightarrow}{b}_2 , \qquad
\stackrel{\rightarrow}{b}_3^* =\stackrel{\rightarrow}{b}_3 .
\end{equation}
Therefore the general solution of Eq.(20) can be written as a linear 
combination \begin{equation}
\stackrel{\rightarrow}{S}(t) =\sum_{i=1}^3 a_i\stackrel{\rightarrow}{S}_i(t)
\end{equation}
of particular solutions
\begin{equation}
\stackrel{\rightarrow}{S}_i(t) =
\stackrel{\rightarrow}{b}_i(t)\exp\{\varphi_i(t)\} ,
\end{equation}
in which $\;\stackrel{\rightarrow}{b}_i\;$ are given by Eq.(24). The 
coefficients in Eq.(26) are defined by the initial condition for spin from 
Eq.(11), which yields
\begin{equation}
a_i =\stackrel{\rightarrow}{S}_0
\stackrel{\rightarrow}{b}_i^*(0) .
\end{equation}
Substituting Eq.(27) into Eq.(20), we obtain the phase
\begin{equation}
\varphi_i(t)=\int_0^t\left [ \omega_2\alpha_i(t) 
-\stackrel{\rightarrow}{b}_i^*(t)\frac{d}{dt}\stackrel{\rightarrow}{b}_i(t)
\right ] dt .
\end{equation}
From Eq.(29), invoking Eq.(25), we find that
\begin{equation}
\varphi_1^* = -\varphi_1 , \qquad \varphi_2^* =-\varphi_2 , \qquad
\varphi_3 =0 .
\end{equation}

Let us accent that, of course, not each system of equations like (20) can 
be solved exactly. The possibility of obtaining here the exact solution (26) 
is due to the antisymmetry of the matrix $\;\hat A\;$.

\section{Atomic Variables}

At the next step of the method of scale separation [36,37] the 
solution for fast variables is to be substituted into the equations for 
slow variables with time averaging the right--hand side of the latter 
equations. As follows from Eqs.(27) and (29), the fast spin fluctuations are 
described by the effective time--dependent frequencies $\;\varphi_i(t)/t\;$.
Hence, the solution (26) does not have a definite period. Consequently, 
the time averaging is given by the rule
\begin{equation}
\langle F\rangle_t \equiv \lim_{\tau\rightarrow\infty}\frac{1}{\tau}
\int_0^\tau F(t)dt .
\end{equation}
Averaging Eq.(26) according to the rule (31) and using Eq.(16), we find
\begin{equation}
\langle\stackrel{\rightarrow}{S}\rangle_t =\frac{1}{2}\left [ (1+x)S_0^x
+yS_0^y -2zS_0^z\right ]( x \stackrel{\rightarrow}{e}_x +
y\stackrel{\rightarrow}{e}_y -4z\stackrel{\rightarrow}{e}_z) .
\end{equation}

The equation for the guiding centers of atomic variables is obtained by 
time--averaging the right--hand side of Eq.(19) with
$\;\stackrel{\rightarrow}{\xi}\;$ treated as a slow variable. Employing 
Eq.(32), we have
\begin{equation}
\frac{d^2\stackrel{\rightarrow}{r}}{dt^2} =\stackrel{\rightarrow}{F}+
\gamma\stackrel{\rightarrow}{\xi} ,
\end{equation}
with the force
\begin{equation}
\stackrel{\rightarrow}{F}=\frac{\omega_1^2}{2}\left [ (1 +x )S_0^x + yS_0^y
-2zS_0^z\right ] ( x \stackrel{\rightarrow}{e}_x + y
\stackrel{\rightarrow}{e}_y +8z \stackrel{\rightarrow}{e}_z ) .
\end{equation}
This force essentially depends on the initial polarization of spins, as it 
should be in the general case and in contrast to the adiabatic force 
mentioned at the end of Sec.II. Only for one type of initial polarization, 
when $\;S_0^x =S\;$ and $\;S_0^y = S_0^z =0\;$, the force (34) reduces to
the adiabatic one. This type of spin polarization leads to the stationary 
confined motion for which the adiabatic approximation is admissible.

Let us analyze another situation when spins are initially polarized along 
the $\;z\;$--axis, so that
\begin{equation}
S_0^x = 0 , \qquad S_0^y =0 , \qquad S_0^z = S .
\end{equation}
Then the force (34) becomes
\begin{equation}
\stackrel{\rightarrow}{F} = - S\omega_1^2 z (\stackrel{\rightarrow}{r} +
7z \stackrel{\rightarrow}{e}_z ) .
\end{equation}
As is evident, the force (36) is nonadiabatic and, even more, it is 
nonpotential since there exists no potential $\;U\;$ such that
$\;\stackrel{\rightarrow}{F}\;$ would be equal to 
$\;-\stackrel{\rightarrow}{\nabla}U\;$.

With the force (36), Eq. (33) becomes
\begin{equation}
\frac{d^2\stackrel{\rightarrow}{r}}{dt^2} + S\omega_1^2 z
(\stackrel{\rightarrow}{r} +7z \stackrel{\rightarrow}{e}_z) =
\gamma\stackrel{\rightarrow}{\xi} .
\end{equation}
The force $\;\gamma\stackrel{\rightarrow}{\xi}\;$, according to (18), 
originates from interatomic interactions in Eq.(9). Because of the isotropy
of the interaction potential $\;\Phi(|\stackrel{\rightarrow}{r}_i -
\stackrel{\rightarrow}{r}_j|)\;$, the force corresponding to these
interactions may be presented as an isotropic vector, that is, we may write
\begin{equation}
\stackrel{\rightarrow}{\xi} =\xi (\stackrel{\rightarrow}{e}_x +
\stackrel{\rightarrow}{e}_y + \stackrel{\rightarrow}{e}_z ) .
\end{equation}
Expanding Eq.(37) into components, we get for the $\;x\;$--component
\begin{equation}
\frac{d^2x}{dt^2} +S\omega_1^2 zx =\gamma\xi .
\end{equation}
The equation for the $\;y\;$--component is the same as Eq.(39) with the 
replacement $\;x\rightarrow y\;$. Therefore, we shall consider in what 
follows only one of the radial components. For the axial variable, Eq.(37) 
gives \begin{equation}
\frac{d^2z}{dt^2} + 8S\omega_1^2 z^2 =\gamma\xi .
\end{equation}
So, we have to solve the system of nonlinear equations (39) and (40).

The general solutions to Eqs.(39) and (40) can be written in the form
\begin{equation}
x=x_1 +x_2 , \qquad z = z_1 + z_2 ,
\end{equation}
in which $\;x_1\;$ and $\;z_1\;$ are the solutions to the corresponding 
homogeneous equations while $\;x_2\;$ and $\;z_2\;$ are the solutions to 
the nonhomogeneous equations. The homogeneous equations are
$$ \frac{d^2x_1}{dt^2} + S\omega_1^2 z_1x_1 = 0 , $$
\begin{equation}
\frac{d^2z_1}{dt^2} + 8S\omega_1^2 z_1^2 = 0 ,
\end{equation}
with the initial conditions
$$ x_1(0) =x_0 , \qquad \dot x_1(0) = v_0^x , $$
\begin{equation}
z_1(0)=z_0 , \qquad \dot z_1(0) = v_0^z ,
\end{equation}
where the dot means the time derivative. Writing down the nonhomogeneous 
equations, simplifying them a little, taking into account that the collision 
rate is typically an order of magnitude smaller than $\;\omega_1\;$, 
yields
$$ \frac{d^2x_2}{dt^2} + S\omega_1^2 z_1x_2 =\gamma\xi - S\omega_1^2x_1z_2, $$
\begin{equation}
\frac{d^2z_2}{dt^2} + 16 S\omega_1^2z_1z_2 =\gamma\xi ;
\end{equation}
the initial conditions for Eq.(44) being
$$ x_2(0) = 0 , \qquad \dot x_2(0) = 0 , $$
\begin{equation}
z_2(0) = 0 , \qquad \dot z_2(0) = 0 .
\end{equation}

To solve the system of equations (42) and (44), we have, first, to solve 
the equation for $\;z_1\;$, than to substitute $\;z_1\;$ into the equation for
$\;x_1\;$, and to use the found $\;x_1\;$ and $\;z_1\;$ in Eq.(44).

\section{Semi--Confined Motion}

The system of equations in (42) for $\;x_1\;$ and $\;z_1\;$ is a system 
of two nonlinear differential equations of second order. However, these 
equations can be solved exactly.

Integrating once the second equation in (42), we get
\begin{equation}
\left (\frac{dz_1}{dt}\right )^2 =\frac{16}{3} S\omega_1^2 (z_m^3 -z_1^3) ,
\end{equation}
where $\;z_m\;$ is an integration constant which can be found from the 
initial conditions in (43) yielding
\begin{equation}
z_m^3 = z_0^3 +\frac{3(v_0^z)^2}{16S\omega_1^2} .
\end{equation}
Since the left--hand side in (46) is nonnegative, this implies that 
$\;z_m^3\geq z_1^3\;$ if $\;S>0\;$, and $\;z_m^3\leq z_1^3\;$ if $\;S<0\;$,
that is, $\;z_m\;$ is the maximal value of $\;z_1\;$ for $\;S>0\;$ and
$\;z_m\;$ is the minimal value of $\;z_1\;$ for $\;S<0\;$,
$$ z_m=\max_{t}z_1(t), \qquad S > 0 ; $$
\begin{equation}
z_m=\min_{t}z_1(t) , \qquad S < 0 .
\end{equation}

We introduce a function
\begin{equation}
{\cal P} = -\frac{4}{3}S\omega_1^2 z_1 ,
\end{equation}
for which Eq.(46) transforms to
\begin{equation}
\left (\frac{d{\cal P}}{dt}\right )^2 = 4{\cal P}^3 -g_2{\cal P} - g_3 ,
\end{equation}
where
\begin{equation}
g_2=0 , \qquad g_3 =-\frac{256}{27} z_m^3S^3\omega_1^6 .
\end{equation}
Eq.(50) is the Weierstrass equation with the Weierstrass invariants in (51) 
and with the discriminant
$$ \Delta\equiv g_2^3 -27g_3^2 = -27g_3^2 . $$
The solution of the Weierstrass equation (50) is the Weierstrass function 
$\;{\cal P}(t-t_0)\;$, where $\;t_0\;$ is an integration constant. The 
Weierstrass function is an elliptic function, that is, a doubly--periodic 
function which is analytic, except at poles, and which has no singularities
other than poles in the finite part of the complex plane. All properties 
of the Weierstrass functions are perfectly described in Refs. [47,48].

In this way, the solution of Eq.(46), and therefore of Eq.(42), reads
\begin{equation}
z_1(t) = -\frac{3}{4S\omega_1^2}{\cal P}(t-t_0) .
\end{equation}
The integration constant $\;t_0\;$ is to be found from the initial 
condition in (43), which gives
\begin{equation}
3{\cal P}(t_0) = -4z_0S\omega_1^2 ,
\end{equation}
where we took into account that the Weierstrass function is an even 
function. When $\;t\;$ tends to $\;t_0\;$, then
\begin{equation}
z_1(t) \simeq -\frac{3}{4S\omega_1^2}\left [\frac{1}{(t-t_0)^2} +
\frac{g_3}{28}(t-t_0)^4\right ] .
\end{equation}
The general behaviour of $\;z_1(t)\;$ is as follows. If $\;S>0\;$, then, 
starting from $\;z_0\;$, the value of $\;z_1(t)\;$ increases to $\;z_m\;$ 
from (47), after which it decreases, diverging to $\;-\infty\;$ as 
$\;t\rightarrow t_0\;$. If $\;S<0\;$, then $\;z_1(t)\;$ decreases to its 
minimal value $\;z_m\;$, after which it turns to the positive direction, 
increasing to $\;+\infty\;$ as $\;t\rightarrow t_0\;$.

Substituting solution (52) into the first of Eqs.(42), we have the equation
\begin{equation}
\frac{d^2x_1}{dt^2} =\frac{3}{4}{\cal P}(t-t_0)x_1
\end{equation}
for the radial motion. This is a Lam\'e equation of degree $\;n=1/2\;$, which
is defined by the relation $\;n(n+1)=3/4\;$. The solution to the Lam\'e 
equation is given by combinations of Lam\'e functions of different kinds 
[49]. In the present case, the solution to Eq.(55) is
\begin{equation}
x_1(t)=\left [ c_1{\cal P}\left (\frac{t-t_0}{2}\right ) + c_2\right ]
E_3^{-1/2}\left (\frac{t-t_0}{2}\right ) .
\end{equation}
Here
$$ E_3(t) \equiv \frac{d}{dt}{\cal P}(t) $$
is a Lam\'e function of degree $\;3\;$, of the first kind. The 
integration constants $\;c_1\;$ and $\;c_2\;$ are defined by the initial 
conditions in (43).

The solution (56) diverges together with (52), as $\;t\rightarrow t_0\;$, 
by the law
\begin{equation}
x_1(t)\simeq c_1\left (\frac{t-t_0}{2}\right )^{-1/2} + c_2\left (
\frac{t-t_0}{2}\right )^{3/2} .
\end{equation}
Comparing Eqs.(57) and (54), we see that the divergence along the axial 
direction is faster than in the radial one; the aspect ratio being
\begin{equation}
\sqrt{\frac{x_1^2(t)}{z_1^2(t)}}\sim |t-t_0|^{3/2} .
\end{equation}
This ratio tends to zero, as $\;t\rightarrow t_0\;$. Therefore, a cloud of 
atoms acquires an ellipsoidal shape stretched in the axial direction.

The behaviour of $\;x_1(t)\;$ and $\;z_1(t)\;$ shows that the atoms with 
an initial polarization $\;S>0\;$ are confined from the side $\;z>0\;$ but
are not confined from the side $\;z<0\;$. Vice versa, the atoms with an 
initial polarization $\;S<0\;$ are confined from the side $\;z<0\;$ but 
are not confined from that $\;z>0\;$. Thus, as ensemble of atoms, with a 
given polarization, loaded into a trap would move predominantly in one 
direction either to $\;z<0\;$ or to $\;z>0\;$ depending on whether the 
initial polarization is $\;S>0\;$ or $\;S<0\;$, respectively. Such a 
regime is exactly that semi--confining regime we have been looking for.

However, we need yet to find the solutions of nonhomogeneous equations in 
(44). To this end, we have to concretize the variable $\;\xi\;$ 
originating from the averaged pair interactions. For a rarefied system these
interactions can be treated as random pair collisions. The system is said 
to be rarefied if the average atomic density $\;\rho\;$ and the 
scattering length $\;a\;$ satisfy the inequality $\;\rho a^3\ll 1\;$. This is
just the case of experiments [5-11] with alkali atoms. Therefore, we can 
consider $\;\xi\;$ as a random variable. More accurately, $\;\xi(t)\;$
can be modelled by a stochastic field. The latter may be specified as 
white Gaussian noise [50] with the stochastic averages
\begin{equation}
\langle\langle\xi(t)\rangle\rangle = 0 , \qquad
\langle\langle\xi(t)\xi(t')\rangle\rangle = 2D\delta(t-t') ,
\end{equation}
where $\;D\;$ is a diffusion rate. Then the nonhomogeneous equations in 
(44) become stochastic differential equations. An explicit way of 
treating the random variable $\xi$ is presented in Appendix.

When solving the nonhomogeneous equations, it is useful to invoke again 
the idea of scale separation. The time variation of solutions to 
stochastic differential equations reflects the properties of the given 
stochastic fields. Since in our case, the stochastic field $\;\xi(t)\;$ is 
modelled by white noise, which is characterized by sharp time jumps 
[50], then the related functional variables $\;x_2\;$ 
and $\;z_2\;$ can be treated as fast compared to $\;x_1\;$ and $\;z_1\;$ 
satisfying the equations not containing such random fields. Hence, for the 
stochastic differential equations in (44), the slow variables $\;x_1\;$ 
and $\;z_1\;$ can be kept as quasi--integrals of motion [36,37]. Then 
using the method of Laplace transforms, we obtain
\begin{equation}
x_2(t) =\int_0^tG_x(t-\tau)\left [\gamma\xi(\tau) - S\omega_1^2x_1z_2(\tau)
\right ]d\tau
\end{equation}
and
\begin{equation}
z_2(t) =\int_0^tG_z(t-\tau)\gamma\xi(\tau)d\tau ,
\end{equation}
where the initial conditions in (45) have been taken into account, and 
the transfer functions are
$$ G_x(t) =\frac{\sin(\varepsilon t)}{\varepsilon} , \qquad
G_z(t) =\frac{\sin(4\varepsilon t)}{4\varepsilon} , $$
with the effective frequency
$$ \varepsilon\equiv\sqrt{Sz_1}\omega_1 . $$

Employing condition (59), we may find the moments of solutions (60) and 
(61). For instance, 
\begin{equation}
\langle\langle x_2(t)\rangle\rangle = 0 , \qquad
\langle\langle z_2(t)\rangle\rangle = 0  .
\end{equation}
Calculating the second moments, we get the mean--square deviations for 
the radial random variable,
$$ \langle\langle x_2^2(t)\rangle\rangle = \frac{\gamma^2Dt}{\varepsilon^2}
\left [ 1 -\frac{\sin(2\varepsilon t)}{2\varepsilon t}\right ] + $$
\begin{equation}
+\frac{\gamma^2Dtx_1^2}{3600\varepsilon^2z_1^2}\left \{ 1 -\cos(\varepsilon t)
\cos(4\varepsilon t) + \frac{\sin(4\varepsilon t)}{4\varepsilon t}\left [
\cos(\varepsilon t) - \cos(4\varepsilon t) -
16\varepsilon t\sin(\varepsilon t)\right ]\right \} ,
\end{equation}
and for the axial random variable,
\begin{equation}
\langle\langle z_2^2(t)\rangle\rangle =\frac{\gamma^2Dt}{16\varepsilon^2}
\left [ 1 -\frac{\sin(8\varepsilon t)}{8\varepsilon t}\right ] .
\end{equation}
Note that the collision rate $\;\gamma\;$ enters here as $\;\gamma^2\;$, 
thence the behaviour of the random variables (63) and (64) does not 
depend on whether the interatomic interactions are repulsive or attractive.

When the variables $\;x_1\;$ and $\;z_1\;$ diverge as in Eqs.(56) and (54), 
then for the random variables in (63) and (64) we obtain
$$ \langle\langle x_2^2(t)\rangle\rangle 
=\propto |t-t_0|^6\exp\left (\frac{4\sqrt{3}t}{|t-t_0|}\right ) , $$
\begin{equation}
\langle\langle z_2^2(t)\rangle\rangle
=\propto |t-t_0|^3\exp\left (\frac{4\sqrt{3}t}{|t-t_0|}\right ) ,
\end{equation}
as $\;t\rightarrow t_0\;$. Since this expansion is governed by the same 
exponentials, it is practically isotropic, with only a slight anisotropy 
due to different preexponential factors.

Remember that the general solutions to the equations (39) and (40) have 
the form of the sums in (41) containing both the regular terms $\;x_1\;$ 
and $\;z_1\;$ and the random terms $\;x_2\;$ and $\;z_2\;$. The relative 
contribution of these terms is regulated by the relation between 
the parameters $\;\gamma, \; D\;$, and $\;\omega_1\;$. If $\;\gamma^2D\ll
\omega_1^3\;$, then the influence of the random terms $\;x_2\;$ and 
$\;z_2\;$ is negligibly small, and the atomic motion is characterized by 
the regular terms $\;x_1\;$ and $\;z_1\;$. In this case we have the 
semi--confining regime. An ensemble of atoms would form an ellipsoidal 
cloud moving in one of the directions along the $\;z\;$--axis. If 
$\;\gamma^2D\gg\omega_1^3\;$, then the motion of atoms is governed by the 
random terms $\;x_2\;$ and $\;z_2\;$. In such a case, an ensemble of 
atoms would form an almost isotropic exponentially expanding cloud.

Assuming, as usual, that the diffusion rate $\;D\;$ is proportional to 
temperature $\;T\;$, we come to the conclusion that the realization 
of either the regime of the fast exponential expansion or the regime of 
the slow semi--confined motion depends on temperature. At high 
temperatures the former regime will be realized while at low 
temperatures, the latter. A crossover temperature $\;T_c\;$ related to 
the equality $\;\gamma^2D=\omega_1^3\;$ would correspond to the effective 
boundary between these two regimes.

\section{Numerical Estimates}

In order to impart to the whole consideration a completely realistic flavour
and to show the reasonability of all inequalities assumed for employing 
the method of scale separation, let us adduce numerical estimates basing 
on the characteristic  quantities typical of the experiments [5-7] with 
$\;^{87}Rb\;$ in dynamical quadrupole traps.

The mass of a Rubidium atom is $\;m=1.45\times 10^{-22}g\;$ and the 
magnetic moment is $\;\mu=0.45\times 10^{-20}erg/G\;$. The gradient of 
the quadrupole field is $\;B_1'=120\; G/cm\;$, and the amplitude of the 
rotating bias field is $\;B_2=10\;G\;$; the rotation frequency of the 
latter being $\;\omega\approx 5\times 10^4s^{-1}\;$. The nonuniformity length
(15) is $\;L\sim 0.1\; cm\;$. The characteristic frequency of atomic 
motion and the Larmor frequency of spins from (17) are $\;\omega_1\sim 
10^2s^{-1}\;$ and $\;\omega_2\sim 5\times 10^7s^{-1}\;$, respectively. The
collision rate in (18) can be estimated as $\;\gamma\sim\hbar\rho a/m\;$,
which, with the average density $\;\rho\approx 3\times 10^{12}cm^{-3}\;$ 
and the scattering length $\;a\sim 10^{-6}cm\;$, gives 
$\;\gamma\sim 10\;s^{-1}\;$. Thus, the following inequalities hold true:
$$ \gamma\ll\omega_1\ll\omega\ll\omega_2 . $$
Hence, the classification of atomic variables as slow and of spin 
variables as fast, based on the inequalities in (21), is correct.

To estimate the lifetime $\;t_0\;$ of the semi--confined motion 
described in Sec.V, we have to return to Eq.(50). 
Integrating this over time between $\;t=0\;$ and some $\;t\;$, we have
$$\int_{{\cal P}(t_0)}^{{\cal P}(t-t_0)}
\frac{d{\cal P}}{\sqrt{4{\cal P}^3-g_2{\cal P} -g_3}} = t . $$
From here, taking into account that $\;g_2=0\;$ and $\;{\cal P}(t-t_0)
\rightarrow\infty\;$, as $\;t\rightarrow t_0\;$, we get
\begin{equation}
t_0=\int_{{\cal P}(t_0)}^{\infty}\frac{d{\cal P}}{\sqrt{4{\cal P}^3-g_3}} .
\end{equation}
Comparing Eqs.(51) and (53), we find for the Weierstrass invariant
$$ g_3 =4\left (\frac{z_m}{z_0}\right )^3{\cal P}^3(t_0) . $$
Then Eq.(66) can be rewritten as
\begin{equation}
t_0 =\tau_0\int_{-\infty}^{z_0}\frac{dz}{\sqrt{z_m^3-z^3}} ,
\end{equation}
where we assume that $S>0$ and 
$$ \tau_0\equiv\frac{1}{4\omega_1}\sqrt{\frac{3}{S}} . $$
The value of $\;z_m\;$ is given by Eq.(47). The average kinetic energy 
$\;\frac{1}{2}m(v_0^z)^2L^2\;$ can be expressed through temperature as 
$\;\frac{1}{2}k_BT\;$. Therefore Eq.(47) acquires the form
\begin{equation}
z_m^3=z_0^3+\zeta ,
\end{equation}
in which
$$ \zeta\equiv\frac{3T}{16T_0}, \qquad 
T_0\equiv\frac{mS\omega_1^2L^2}{k_B}\; . $$
Finally, for the lifetime (67) we obtain
\begin{equation}
t_0 =\tau_0\int_{-z_0}^{\infty}\frac{dz}{\sqrt{z^3+z_0^3+\zeta}} .
\end{equation}
For the quantities considered, we have $\;\tau_0\approx 10^{-2}s\;$ and 
$\;T_0\approx 10^{-4}K\;$. The parameter $\;\zeta\;$ depends on 
temperature. In the interval of temperatures between $\;1\;nK\;$ and 
$\;1\;mK\;$, it changes from $\;10^{-6}\;$ to $\;1\;$. The quantity $z_0$, 
according to Eq.(43), is the initial coordinate $z_1(0)$. It is clear 
that a cloud of trapped atoms should have a distribution of $z_0$ which 
depends on the temperature and shape of the confining potential. The 
average value of $z_0$ can be interpreted as the initial location of the 
center of an atomic cloud. This value can be different for different 
experiments. Keeping this in mind, we consider below several values of $z_0$.
The integral (69) was calculated numerically for different initial positions 
$\;z_0\;$. The results are presented in Table 1. As is seen, for the 
temperatures $\;T\approx 10^{-5}K\;$ the time $\;t_0\;$ becomes of order 
$\;0.1\;s\;$, and for $\;T\approx 1\; nK\;$ it reaches 
$\;t_0\approx 0.3\; s\;$. Moreover, diminishing the temperature lower, it is 
possible to make $\;t_0\;$ arbitrary large, since $\;t_0\rightarrow\infty\;$ 
when $\;\zeta\rightarrow 0\;$. Notice that the lifetime $t_0$ does not 
change much when varying $z_0$ in the interval $-0.1\leq z_0\leq 0.1$.

It is possible to pose the question: how well does the lifetime $\;t_0\;$ 
characterize the real escape time of an atom from a trap? Looking at 
Eq.(54), we see that $\;z_1(t)\;$ diverges as $\;t\rightarrow t_0\;$, while, 
according to condition (16), the actual escape of an atom from a trap 
occurs at the time $\;t_1\;$, when $\;z(t_1)\sim 1\;$. The relation 
between the times $\;t_0\;$ and $\;t_1\;$, as follows from Eq.(54), is 
$\;t_1-t_0\sim\omega_1^{-1}\;$. This, with the given $\;\omega_1\sim 
10^2s^{-1}\;$, makes $\;t_1-t_0\sim 10^{-2}s\;$. So, if $\;t_0\geq 
0.1\;s\;$, then $\;t_1\sim t_0\;$. Therefore, the time $\;t_0\;$  really 
plays the role of the average lifetime of atoms in a trap during the 
regime of semi--confined motion.

If we are not satisfied by the simple estimates for the 
characteristic time $\;t_1\;$, we can calculate it {\it exactly} from the 
equation of motion. The procedure is the same as that for calculating 
$\;t_0\;$. The results of this calculation confirm that $\;t_1\;$ is very 
close to the time $\;t_0\;$, being smaller by about $\;0.02\;s\;$. 
Because of the mutual closeness of these times, we do not repeat for 
$\;t_1\;$ the whole table as for $\;t_0\;$ but, for cogency, we present in 
Table 2 the values of $\;t_1\;$ for the particles with the initial 
location at the centre of the trap. These values are found from the formula
\begin{equation}
t_1 =\tau_0\int_0^1\frac{dz}{\sqrt{z^3+\zeta}} .
\end{equation}
Similarly to $\;t_0\;$, the time $\;t_1\;$ also becomes arbitrary large, 
$\;t_1\rightarrow\infty\;$, as $\;\zeta\rightarrow 0\;$, i.e. when 
temperature decreases.

Treating the interactions of atoms as random pair collisions is admissible 
if the atomic system is rarefied, so that $\;\rho a^3\ll 1\;$. For the 
case considered, $\;\rho\sim 10^{12}cm^{-3}\;$ and $\;a\sim 10^{-6}cm\;$, 
yielding $\;\rho a^3\sim 10^{-6}\;$. Thus, the assumed inequality is 
well satisfied.

As is argued at the end of Sec.V, at high temperatures there exists a 
regime of fast exponential expansion of an almost isotropic cloud. At low 
temperatures, the regime changes to the semi--confined motion of a slowly 
moving ellipsoidal cloud. The crossover temperature $\;T_c\;$ can be 
defined by the equality $\;\gamma^2D=\omega_1^3\;$. The latter, with the 
diffusion rate $\;D\sim k_BT/\hbar\;$, gives
$$ T_c\sim\frac{\hbar\omega_1^3}{k_B\gamma^2} . $$
Substituting here $\;\omega_1\sim 10^2s^{-1}\;$ and $\;\gamma\sim 
10\;s^{-1}\;$, we have $\;T_c\sim 10^{-7}K\;$, which is close to the 
temperature of the Bose condensation observed in experiments 
[6,7]. Consequently, the semi--confined regime can also be realized under
similar conditions.

Keeping in mind the future possibility of applying statistical methods to 
describe an ensemble of atoms in the semi--confining regime, we have 
to understand whether the local equilibrium can be established in this 
case. The time of local equilibrium is related to the collision rate as 
$\;\tau_{loc}\sim\gamma^{-1}\;$. Hence for $\;\gamma\sim 10\; s^{-1}\;$, we
have $\;\tau_{loc}\sim 0.1\; s\;$. The local equilibrium develops if the 
lifetime of the cloud, $\;t_0\;$, is longer than the local--equilibrium
time. If $\;t_0\sim 0.3\; s\;$, then the local equilibrium can be achieved. 
Of course, the global equilibrium for a semi--confined motion cannot exist. 
Recall that the corresponding force (36) is not confining for a semi--axis of 
$\;z\;$, and furthermore it is not potential. But the possibility of local 
equilibrium implies that a statistical description for such a regime can 
be done, e.g., with the help of hydrodynamic equations. If the local 
equilibrium is absent, one has to use kinetic equations.

\section{Conclusion}

The method of scale separation is applied to the dynamics of 
neutral atoms in nonuniform magnetic fields typical of quadrupole 
magnetic traps. We concentrate our attention on the case of a dynamic 
quadrupole trap with a rotating bias field, as in experiments [5-7]. For 
the initial spin polarization $\;S_0^x<0\;$, the motion of atoms is 
confined and well described by the adiabatic approximation.

For the atoms whose spins are initially polarized along the $\;z\;$--axis a 
novel unusual regime appears where the motion is confined in a half--space: 
when $\;S_0^z>0\;$, the motion is confined from the side $\;z>0\;$, and 
if $\;S_0^z<0\;$, it is confined from the side $\;z<0\;$. This 
semi--confined motion is nonadiabatic and nonpotential. An ensemble of 
atoms in a semi--confining regime forms an ellipsoidal cloud stretched 
in the axial direction and slowly moving along the $\;z\;$-axis to the 
nonconfining side.

The semi--confining regime can be used for studying mixtures with a 
relative motion of components. For example, one component can have 
initial spin polarization $\;S_0^x<0\;$, thus, being confined, while another 
component, with initial spin polarization along the $\;z\;$--axis, 
can be 
semi--confined, moving through the first. Or one can load into a trap 
two components both being polarized along the $\;z\;$--axis, but one with 
$\;S_0^z>0\;$ while another with $\;S_0^z<0\;$. Then both such species 
will be in the semi--confining regime but moving in opposite directions, 
one to the negative $\;z\;$-- and another to positive $\;z\;$--direction. 
The variant with three components, one of which is confined with the two 
others semi--confined moving in opposite directions, could also be realized. 
Mixtures with the relative motion of components can display a number 
of interesting features, e.g., the effect of conical stratification [33].

The semi--confining regime can also be employed for separating the species 
with different initial polarizations. For instance, if one loads into 
a trap a mixture of two components, one having the spin polarization up 
and another down, then the trap will act as a separator moving the first 
component down and the second one up, thus separating them.

One more application of the semi--confining regime could be for realizing 
atom lasers, for which a directed motion of atoms is necessary.

\vspace{5mm}

{\bf Acknowledgements}

\vspace{2mm}

I am very grateful to V.S. Bagnato and E.A. Cornell for consultations on 
experiments with trapped alkali atoms and to E.P. Yukalova for mathematical 
advice and numerical calculations. 

Financial support of the National Science and Technology Development Council
of Brazil is appreciated.

\vspace{1cm}

{\Large{\bf Appendix. Random Variables}}

\vspace{0.5cm}

The variable $\xi(t)$ describing random pair collisions of atoms has been 
treated as a stochastic variable. Consequently, Eqs.(44) are stochastic 
differential equations. To find their solution, it was necessary to 
define  the stochastic averages (59). The explicit definition of the 
random variable $\xi(t)$ and of the corresponding stochastic averages can 
be done in the following way.

Assume that there is a set $\{\varphi_n(t)\}$ of functions $\varphi_n(t)$ 
ennumerated by a multi--index $n$. Let this set be complete and orthonormal,
$$ \sum_n\varphi_n^*(t)\varphi_n(t')=\delta(t-t') , \qquad
\int\varphi_m^*(t)\varphi_n(t)dt =\delta_{mn} . $$
Then a function $\xi(t)$ can be represented as an expansion
$$ \xi(t) = \sum_n\xi_n\varphi_n(t) . $$
Each coefficient $\xi_n$ is considered as a random variable with a probability
distribution $p(\xi_n)$. For concreteness, we may think of $p_n(\xi_n)$ 
as a Gaussian distribution.

The stochastic averaging for a function $F(\xi(t))$ of the stochastic 
variable $\xi(t)$ is defined as the functional integral
$$ \langle\langle F(\xi(t))\rangle\rangle = \int F(\xi(t))\prod_n
p_n(\xi_n)d\xi_n . $$
If the random coefficient $\xi_n$ is complex, then $d\xi_n\equiv 
d({\rm Re}\xi_n)d({\rm Im}\xi_n)$. When $\xi_n$ is centered at zero, then
$$ \langle\langle\xi_n\rangle\rangle =\int\xi_np_n(\xi_n)d\xi_n = 0 . $$
The dispersion $\sigma_n$ of a distribution $p_n(\xi_n)$ is given by the 
equation
$$ \sigma_n^2\equiv
\langle\langle|\xi_n|^2\rangle\rangle =\int|\xi_n|^2p_n(\xi_n)d\xi_n . $$

With these definitions, for the stochastic correlation function we have
$$  \langle\langle\xi^*(t)\xi(t')\rangle\rangle =\sum_n
\sigma_n^2\varphi_n^*(t)\varphi_n(t') . $$
In the case of white noise, all dispersions $\sigma_n$ are equal to each 
other, so that we may write $\sigma_n^2=2D$. Thence, the stochastic 
correlation function becomes
$$ \langle\langle\xi(t)\xi(t')\rangle\rangle = 2D\sum_n
\varphi_n^*(t)\varphi_n(t') , $$
where it is taken into account that $\xi(t)$ is real. From here, because 
of the completeness of the basis $\{\varphi_n(t)\}$, we obtain Eq.(59) 
which was used in calculating expressions (62)--(65).

\newpage

\begin{center}
{\bf Table Captions}
\end{center}

{\bf Table 1.}

\vspace{3mm}

The characteristic time $\;t_0\;$, in seconds, for several parameters 
$\;\zeta\;$ and different initial conditions.

\vspace{1cm}

{\bf Table 2.}

\vspace{3mm}

The characteristic time $\;t_1\;$, in seconds, for the atoms initially 
located at the centre of the trap and for different parameters $\;\zeta\;$.

\newpage

\begin{center}

{\bf Table 1}

\vspace{5mm}

\begin{tabular}{|c|c|c|c|c|c|c|c|}\hline
$\zeta$&$10^{-6}$&$10^{-5}$&$10^{-4}$&$10^{-3}$&$10^{-2}$&$10^{-1}$&$1$ \\
$z_0$  &         &         &         &         &         &         & \\ \hline
-0.1   &  0.08   &  0.07   &  0.07   &  0.06   &  0.05   &  0.04   & 0.03 \\
-0.01  &  0.20   &  0.16   &  0.12   &  0.09   &  0.06   &  0.04   & 0.03 \\
-0.001 &  0.27   &  0.19   &  0.13   &  0.09   &  0.06   &  0.04   & 0.03 \\
0      &  0.28   &  0.19   &  0.13   &  0.09   &  0.06   &  0.04   & 0.03 \\
 0.001 &  0.29   &  0.19   &  0.13   &  0.09   &  0.06   &  0.04   & 0.03 \\
 0.01  &  0.33   &  0.22   &  0.14   &  0.09   &  0.06   &  0.04   & 0.03 \\
 0.1   &  0.13   &  0.13   &  0.12   &  0.10   &  0.07   &  0.04   & 0.03 \\
\hline
\end{tabular}

\vspace{3cm}

{\bf Table 2}

\vspace{5mm}

\begin{tabular}{|c|c|c|c|c|c|c|c|}\hline

$\zeta$ & $10^{-6}$&$10^{-5}$&$10^{-4}$&$10^{-3}$&$10^{-2}$&$10^{-1}$&$1$\\
\hline
$t_1$  & 0.26     & 0.17    & 0.11    & 0.07    &  0.04   & 0.02   & 0.01\\
\hline
\end{tabular}

\end{center}


\newpage

\begin{references}
\bibitem{1} D.E. Pritchard, Phys. Rev. Lett. {\bf 51}, 1336 (1983).

\bibitem{2} H.F. Hess, Phys. Rev. B {\bf 34}, 3476 (1986).

\bibitem{3} V.S. Bagnato, G.P. Lafyatis, A.G. Martin, E.L. Raab, R.N. 
Ahmad--Bitar, and D.E. Pritchard, Phys. Rev. Lett. {\bf 58}, 2194 (1987).

\bibitem{4} E.A. Cornell, C. Monroe, and C.E. Wieman, Phys. Rev. Lett. 
{\bf 67}, 2439 (1991).

\bibitem{5} W. Petrich, M.H. Anderson, J.R.Ensher, and E.A. Cornell, 
Phys. Rev. Lett. {\bf 74}, 3352 (1995).

\bibitem{6} M.H. Anderson, J.R. Ensher, M.R. Matthews, C.E. Wieman, and 
E.A. Cornell, Science {\bf 269}, 198 (1995).

\bibitem{7} D.S. Jin, J.R. Ensher, M.R. Matthews, C.E. Wieman, and E.A. 
Cornell, Phys. Rev. Lett. {\bf 77}, 420 (1996).

\bibitem{8} C.C. Bradley, C.A. Sackett, J.J. Tollet, and R.G. Hulet, 
Phys. Rev. Lett. {\bf 75}, 1687 (1995).

\bibitem{9} K.B. Davis, M.O. Mewes, M.R. Andrews, N.J. van Druten, D.S. 
Durfee, D.M. Kurn, and W. Ketterle, Phys. Rev. Lett. {\bf 75}, 3969 (1995).

\bibitem{10} M.O. Mewes, M.R. Andrews, N.J. van Druten, D.M. Kurn, D.S. 
Durfee, and W. Ketterle, Phys. Rev. Lett. {\bf 77}, 416 (1996).

\bibitem{11} M.O. Mewes, M.R. Andrews, N.J. van Druten, D.M. Kurn, D.S. 
Durfee, C.G. Townsend, and W. Ketterle, Phys. Rev. Lett. {\bf 77}, 988 
(1996).

\bibitem{12} V. Bagnato, D. Pritchard, and D. Kleppner, Phys. Rev. A {\bf 
35}, 4354 (1987).

\bibitem{13} V.S. Bagnato, Phys. Rev. A {\bf 54}, 1726 (1996).

\bibitem{14} M. Holland and J. Cooper, Phys. Rev. A {\bf 53}, 1954 (1996).

\bibitem{15} F. Dalfovo and S. Stringari, Phys. Rev. A {\bf 53}, 2477 (1996).

\bibitem{16} M. Edwards, P.A. Ruprecht, K. Burnett, R.J. Dodd, and C.W. 
Clark, Phys. Rev. Lett. {\bf 77}, 1671 (1996).

\bibitem{17} W. Krauth, Phys. Rev. Lett. {\bf 77}, 3695 (1996).

\bibitem{18} J. Oliva, Phys. Rev. B {\bf 39}, 4197 (1989).

\bibitem{19} T.L. Ho and V.B. Shenoy, Phys. Rev. Lett. {\bf 77}, 3276 (1996).

\bibitem{20} K.G. Singh and D.S. Rokhsar, Phys. Rev. Lett. {\bf 77}, 1667
(1996).

\bibitem{21} F. Brosens, J.T. Devreese, and L.F. Lemmens, Solid State Commun.
{\bf 100}, 123 (1996).

\bibitem{22} K. Kirsten and D.J. Toms, Phys. Rev. A {\bf 54}, 4188 (1996).

\bibitem{23} J. Oliva, Phys. Rev. B {\bf 39}, 4204 (1989).

\bibitem{24} D.A. Butts and D.S. Rokhsar, quant-ph/ 9612027 (1996).

\bibitem{25} A.L. Migdall, J.V. Prodan, W.D. Phillips, T.H. Bergeman, and 
H.J. Metcalf, Phys. Rev. Lett. {\bf 54}, 2596 (1985). 

\bibitem{26} T. Bergeman, G. Erez, and H. Metcalf, Phys. Rev. A {\bf 35}, 
1353 (1987).

\bibitem{27} T.H. Bergeman, P. McNicholl, J. Kyacia, H. Metcalf, and N.L. 
Balazs, J. Opt. Soc. Am. B {\bf 6}, 2249 (1989).

\bibitem{28} A. Gongora--T., A. Antillon, and T.H. Seligman, Phys. Rev. A 
{\bf 42}, 3139 (1990).

\bibitem{29} A. Antillon, A. Gongora--T., and T.H. Seligman, Z. Phys. D 
{\bf 24}, 347 (1992).

\bibitem{30} W. S\"uptitz, G. Wokurka, F. Strauch, P. Kohns, and W. 
Ertmer, Opt. Lett. {\bf 19}, 1571 (1994).

\bibitem{31} M.S. Santos, P. Nussenzveig, L.G. Marcassa, K. Helmerson, J. 
Flemming, S.C. Zilio, and V.S. Bagnato, Phys. Rev. A {\bf 52}, 4340 (1995).

\bibitem{32} H.T. Tan, C.W. Woo, and F.Y. Wu, J. Low Temp. Phys. {\bf 5}, 
261 (1971).

\bibitem{33} V.I. Yukalov, Acta Phys. Pol. A {\bf 57}, 295 (1980).

\bibitem{34} Y.A. Nepomnyashchy, Phys. Rev. B {\bf 47} 905 (1993).

\bibitem{35} L.D. Landau and E.M. Lifshitz, {\it Quantum Mechanics} 
(Pergamon, Oxford, 1987).

\bibitem{36} V.I. Yukalov, Phys. Rev. Lett. {\bf 75}, 3000 (1995).

\bibitem{37} V.I. Yukalov, Phys. Rev. B {\bf 53}, 9232 (1996).

\bibitem{38} N.N. Bogolubov and Y.A. Mitropolsky, {\it Asymptotic Methods 
in Theory of Nonlinear Oscillations} (Gordon and Breach, New York, 1961).

\bibitem{39} H. Poincar\'e, {\it New Methods of Celestial Mechanics}
(AIP, New York, 1993).

\bibitem{40} V.I. Yukalov, Laser Phys. {\bf 5}, 526 (1995).

\bibitem{41} V.I. Yukalov, Laser Phys. {\bf 5}, 970 (1995).

\bibitem{42} V.I. Yukalov, Nucl. Instrum. Methods Phys. Res. A {\bf 370}, 
345 (1996).

\bibitem{43} J.F. Kiselev, A.F. Prudkoglyad, A.S. Shumovsky, and V.I. 
Yukalov, Mod. Phys. Lett. B {\bf 1}, 409 (1988). 

\bibitem{44} Y.F. Kiselev, A.F. Prudkoglyad, A.S. Shumovsky, and V.I. 
Yukalov, J. Exp. Theor. Phys. {\bf 67}, 413 (1988). 

\bibitem{45} T.S. Belozerova, V.K. Henner, and V.I. Yukalov, Comput. 
Phys. Commun. {\bf 73}, 151 (1992).

\bibitem{46} T.S. Belozerova, V.K. Henner, and V.I. Yukalov, Phys. Rev. B
{\bf 46}, 682 (1992).

\bibitem{47} H. Bateman and A. Erd\'elyi, {\it Higher Transcendental 
Functions}, Vol. 2 (Krieger, Malabar, 1985).

\bibitem{48} E.T. Whittaker and G.N. Watson, {\it Modern Analysis} 
(Cambridge Univ., Cambridge, 1988).

\bibitem{49} H. Bateman and A. Erd\'elyi, {\it Higher Transcendental 
Functions}, Vol. 3 (Krieger, Malabar, 1987).

\bibitem{50} J. Honerkamp, {\it Stochastic Dynamical Systems} (VCH, New 
York, 1994).
\end{references}
\end{document}